\documentstyle[pra,aps,twocolumn]{revtex}

\def\QED{\leavevmode\unskip\penalty9999 \hbox{}\nobreak\hfill
     \quad\hbox{\leavevmode  \hbox to.77778em{%
               \hfil\vrule   \vbox to.675em%
               {\hrule width.6em\vfil\hrule}\vrule\hfil}}
     \par\vskip24pt}

\def\ibb #1{\leavevmode\hbox{\kern.3em\vrule
     height 1.5ex depth -.1ex width .4pt\kern-.3em\rm#1}}
\def\Cx {{\ibb C}}

\newtheorem{Def}{Definition}
\newtheorem{Lem}{Proposition}

\title{Bound entangled Gaussian states}
 \author{R.~F. Werner\thanks{Electronic Mail: \tt{r.werner@tu-bs.de}}
 {{}\ and\ }M.~M. Wolf\thanks{Electronic Mail: \tt{mm.wolf@tu-bs.de}}
   \\[1ex]
  {\small Institut f{\"u}r Mathematische Physik, TU Braunschweig,}\\
  {\small Mendelssohnstr.3, 38106 Braunschweig, Germany.}}
\date{\today}

\begin{document}
\draft \maketitle

\begin{abstract}We discuss the entanglement properties of bipartite
states with Gaussian Wigner functions. Separability and the
positivity of the partial transpose are characterized in terms of
the covariance matrix of the state, and it is shown that for
systems composed of a single oscillator for Alice and an arbitrary
number for Bob, positivity of the partial transpose implies
separability. However, this implications fails with two
oscillators on each side, as we show by a five parameter family of
explicit counterexamples.
\end{abstract}

\pacs{03.65.Bz, 03.67.-a}

\narrowtext

\section{Introduction}

Many experiments in the young field of quantum information physics
are not carried out on finite dimensional quantum systems, for
which most of the basic theory has been developed, but in the
quantum optical setting. In that setting the basic variables are
quadratures of field modes, which satisfy canonical commutation
relations, and hence have no finite dimensional realizations. It
would seem that the theory therefore becomes burdened with all the
technical difficulties of infinite dimensional spaces, while
theoreticians are on the other hand still struggling to answer
some simple questions about qubit systems. However, the states
relevant in quantum optics are often of a special kind, and for
this class the typical questions of quantum information theory are
luckily  of the same  complexity as for the usual finite
dimensional systems.

This simple class of states of ``continuous variable systems'' is
the class of {\it Gaussian states}, i.e., those states whose
Wigner function is a Gaussian on phase space. Such a state is
therefore completely specified by its mean and its covariance
matrix, where the mean is irrelevant for entanglement questions,
because it can be shifted to zero by a local unitary (phase space
translation). It turns out that the basic entanglement properties
of a Gaussian density matrix (as a state on two infinite
dimensional Hilbert spaces) can be translated very nicely into
properties of its covariance matrix (see Section~2), so that
problems involving Gaussian states are reduced to problems of
finite dimensional linear algebra rather reminiscent of the
problems involving finite dimensional density matrices.

For the latter it is well known \cite{Peres,Hor1} that the
positivity of the partial transpose (``ppt'') is necessary for
separability, but sufficient only for the smallest non-trivial
systems, namely systems in dimensions $2\otimes2$ and $2\otimes3$.
In all higher dimensions we can find ``bound entangled states'',
which are not separable, but nevertheless have a positive partial
transpose, and are hence not distillible \cite{Hor2}. In the case
of continuous variable systems the first nontrivial examples of
this kind were obtained in \cite{HorLev}. In the Gaussian setting
it was shown by Simon \cite{Simon} that for bipartite systems with
one canonical degree of freedom on each side (Alice and Bob),
i.e., once again for the simplest possible systems, the
equivalence of ppt and separability also holds. For this system it
was also shown that non-ppt states are indeed distillible
\cite{Zoller}. In the present paper we settle the relationship
between separability and ppt for all higher dimensions, showing
that the equivalence holds also for systems of $1\times N$
oscillators, but fails for all higher dimensions. We show this by
giving explicit examples for $2\times2$ oscillators.

The key idea for constructing bound entangled Gaussian states is
the notion of ``minimally ppt'' covariance matrices. These are
defined as the covariance matrices of ppt Gaussian states, which
are not larger (in matrix ordering) than the covariance matrix of
any other ppt Gaussian state. It is easy to see that a minimally
ppt covariance matrix belongs to a separable state iff that state
is a product state. Hence bound entangled Gaussians arise from all
minimally ppt covariance matrices, which are not block diagonal.
Numerically, minimally ppt covariance matrices can be obtained
very efficiently by successively subtracting rank one operators
from a given covariance matrix. This algorithm is reminiscent of
techniques for density matrices in the context of ``best separable
approximation'' \cite{BSA}. Running this procedure for $2\times2$
or larger systems generically gives bound entangled Gaussian
states.

Our paper is organized as follows: In Section~2 we will set up the
basic notation, and the translation of separability and ppt
conditions into properties of covariance matrices (for
separability this appears to be new). We also describe the
minimally ppt covariance matrices. In Section~3 we prove the
equivalence for the $1\times N$ case, and in Section~4 we present
a five parameter family of $2\times2$ bound entangled states.
\section{Gaussian states and entanglement}
A system of $f$ canonical degrees of freedom is described
classically in a phase space, which is a  $2f$-dimensional real
vector space $X$. The canonical structure is given by a
$2f\times2f$ matrix $\sigma$, known as the {\it symplectic
matrix}, which is antisymmetric and non-singular. With a suitable
choice of coordinates (``canonical coordinates''), it can be
brought into a standard form:  The $2f$ variables are then grouped
into $f$ canonical pairs (e.g., position and momentum), for each
of which the symplectic matrix takes the form
$\sigma=\bigl(\begin{array}{cc}0&-1\\1&0\end{array}\bigr)$, and
all other matrix elements vanish.

The symplectic matrix also governs the canonical commutation
relations for the corresponding quantum system: if $R_\alpha$,
$\alpha=1,\ldots,2f$ are canonical operators (for canonical
coordinates these are naturally grouped into $f$ standard position
operators and $f$ standard momentum operators), the commutation
relations read
\begin{equation}\label{ccr}
  i[R_\alpha,R_\beta]=\sigma_{\alpha\beta}{\bf 1}\;.
\end{equation}
These relations may be exponentiated to the Weyl relations
involving unitaries $W(\xi)=\exp\bigl(i\xi\cdot\sigma\cdot R)$,
where $\xi\in X$, and $\xi\cdot\sigma\cdot
R=\sum_{\alpha\beta}\xi_\alpha\sigma_{\alpha\beta}R_\beta$. These
{\it Weyl operators} implement the phase space translations. We
will assume that they act irreducibly on the given Hilbert space,
i.e., that there are no further degrees of freedom. Then by von
Neumann's uniqueness Theorem \cite{vN} the $R_\alpha$ are
unitarily equivalent to the usual position and momentum operators
in the ${\cal L}^2$ space over position space.

For a general density operator $\rho$ we define the {\it mean} as
the vector $m_\alpha={\rm tr}(\rho R_\alpha)$, and the {\it
covariance matrix} $\gamma$ by
\begin{equation}\label{cov}
  \gamma_{\alpha\beta}+i\sigma_{\alpha\beta}
      =2\;{\rm tr}\Bigl[\rho(R_\alpha-m_\alpha{\bf 1})
          (R_\beta-m_\beta{\bf 1})\Bigr]\;,
\end{equation}
which is well-defined whenever all of the unbounded positive
operators $R_\alpha^2$ have finite expectations in $\rho$.  Due to
the canonical commutation relations the antisymmetric part of the
right hand side is indeed the symplectic matrix, independently of
the state $\rho$. The state-dependent covariance matrix $\gamma$
is therefore real and symmetric. Moreover, $\gamma+i\sigma$ is
obviously positive definite.

A Gaussian state is best defined in terms of its {\it
characteristic function}, which for a general state is
$\xi\mapsto{\rm tr}(\rho W(\xi))$. This should be seen as the
quantum Fourier transform \cite{QHA} of $\rho$, and is indeed the
Fourier transform of the Wigner function of $\rho$. Hence we call
$\rho$ {\it Gaussian}, if its characteristic function is of the
form
\begin{equation}\label{charfunc}
 {\rm tr} \bigl[\rho W(\xi)\bigr]=\exp\bigl(i m^T \xi - \frac14\xi^T \gamma
 \xi\bigr)\;.
\end{equation}
Here the coefficients were chosen such that $\gamma$ and $m$ are
indeed covariance and mean of $\rho$, as is readily verified by
differentiation. The necessary condition $\gamma+i\sigma\geq0$,
which is equivalent to $\gamma-i\sigma\geq0$ by complex
conjugation, is also sufficient for Equation~(\ref{charfunc}) to
define a positive operator $\rho$. We note for later use that a
Gaussian state is pure iff $(\sigma^{-1}\gamma)^2=-{\bf
1}$\cite{Hol}, which is equivalent to $\gamma+i\sigma$ having
maximal number of null eigenvectors, i.e., the null space ${\cal
N}=\big\lbrace\Phi\vert(\gamma+i\sigma)\Phi=0\big\rbrace$ has
dimension $(\dim X)/2$. Note that this null space must always be
considered as a complex linear subspace of $\Cx^{2f}$, the
complexification of $X$. For such a complex subspace we denote by
${\Re\!e}{\cal N}$ the subspace of $X$ consisting of all real
parts of vectors in ${\cal N}$. Then a Gaussian state is pure iff
${\Re\!e}{\cal N}=X$.

Let us now consider bipartite systems. The phase space is then
split into two phase spaces $X=X_A\oplus X_B$, where $A$ stands
for Alice and $B$ for Bob. This is a ``symplectic direct sum'',
which means that $\sigma=\sigma_A\oplus\sigma_B$ is block diagonal
with respect to this decomposition. In other words, Alice's
canonical operators $R_\alpha$ commute with all of Bob's. The Weyl
operators are naturally identified with tensor products:
$W(\xi_A\oplus\xi_B)\cong W(\xi_A)\otimes W(\xi_B)$. We call this
an $f_A\times f_B$ system, if $\dim X_A=2f_A$, and $\dim
X_B=2f_B$.

It is clear from (\ref{cov}) and (\ref{charfunc}) that the
covariance matrix of a product state is block diagonal and,
conversely, a Gaussian state with block diagonal $\gamma$ is a
product state. Separability is characterized as follows:

\begin{Lem}\label{L:sep}
Let $\gamma$ be the covariance matrix of a separable state with
finite second moments. Then there are covariance matrices
$\gamma_A$ and $\gamma_B$ such that
\begin{equation}\label{sep}
  \gamma\geq\left(\begin{array}{cc}\gamma_A&0\\0&\gamma_B\end{array}\right)\;.
\end{equation}
Conversely, if this condition is satisfied, the Gaussian state
with covariance $\gamma$ is separable.
\end{Lem}

In order to show the first statement suppose the given state is
decomposed into product states with covariance $\gamma^k$ and mean
$m^k$ with convex weight $\lambda_k$. Then
$m_\alpha=\sum_k\lambda_km_\alpha^k$ and, similarly, for the
second moments we have
\begin{equation}\label{secmom}
  \gamma_{\alpha\beta}+ 2 m_\alpha m_\beta=
     \sum_k\lambda_k\bigl(\gamma^k_{\alpha\beta}
         + 2 m^k_\alpha m^k_\beta\bigr)\;.
\end{equation}
Hence the difference between $\gamma$ and the block diagonal
$\sum_k\lambda_k\gamma^k$ is the matrix
\begin{equation}\label{del}
  \Delta_{\alpha\beta}=
      2 \Bigl( \sum_k\lambda_km^k_\alpha m^k_\beta -
      \sum_{k\ell}\lambda_k\lambda_\ell
          m^k_\alpha m^\ell_\beta\Bigr)\;,
\end{equation}
which is positive definite, because
$\sum\xi_\alpha\xi_\beta\Delta_{\alpha\beta}
  =\sum_{k\ell}\lambda_k\lambda_\ell (s_k-s_\ell)^2\geq0$,
where $s_k=\sum_\alpha\xi_\alpha m_\alpha$.

In order to show the converse, let $\sigma$ be the Gaussian
product state with covariance $\gamma_A\oplus\gamma_B$, and let
$\gamma'=\gamma-\gamma_A\oplus\gamma_B\geq0$. Then $\gamma'$ is
the covariance of a classical Gaussian probability distribution
$P$, and the characteristic function of the given state $\rho$ is
the product of the characteristic function of $\sigma$ and the
Fourier transform of $P$. Hence $\rho$ is the convolution of
$\sigma$ and $P$ in the sense of \cite{QHA}, which is the average
of the phase space translates $W(\xi)\sigma W(\xi)^*$ over $\xi$
with weight $P$. Since all these states will be product states,
$\rho$ is separable. \QED

There are different ways of characterizing the partial transpose.
One simple way is to say that with respect to some set of
canonical coordinates the momenta in Alice's system are reversed,
while her position coordinates and all of Bob's canonical
variables are left unchanged. In addition, the order of factors in
the partial transpose of $R_\alpha R_\beta$ is reversed when both
factors belong to Alice. When we replace $\rho$ in (\ref{cov}) by
its partial transpose, we therefore find the antisymmetric part of
the equation unchanged, whereas $\gamma_{\alpha\beta}$ picks up a
factor $-1$ whenever just one of the indices corresponds to one of
Alice's momenta.  Let us call the resulting covariance matrix by
$\widetilde\gamma$. Clearly, if the partial transpose of $\rho$ is
again a density operator, we must have
$\widetilde\gamma+i\sigma\geq0$. But this is equivalent to
$\gamma+i\widetilde\sigma\geq0$, where in $\widetilde\sigma$ the
corresponding components are reversed, so that
$\widetilde\sigma=(-\sigma_A)\oplus\sigma_B$. This form of the
condition is even valid if we do not insist on canonical
variables. Combining it with the positivity condition for Gaussian
states we get the following characterization:

\begin{Lem}\label{L:ppt} Let $\gamma$ be the covariance matrix of a
state with finite second moments, which has positive partial
transpose. Then
\begin{equation}\label{ppt}
  \gamma+i\widetilde\sigma\geq0 \ ,\ \text{where }\
  \widetilde\sigma=\left(\begin{array}{cc}-\sigma_A&0\\
                  0&\sigma_B\end{array}\right)\;.
\end{equation}
Conversely, if this condition is satisfied, the Gaussian state
with covariance $\gamma$ has positive partial transpose.
\end{Lem}

When $\rho$ is separable, Proposition~\ref{L:sep} shows the
existence of a block diagonal $\gamma'=\gamma_A\oplus\gamma_B$
with $\gamma\geq\gamma'$. Since $\gamma_A$ and $\gamma_B$ are
covariance matrices in their own right, we have $\gamma_A\pm
i\sigma_A\geq0$, and similarly for Bob's side. But this means that
$\gamma\geq\gamma'\geq-i\sigma$ and
$\gamma\geq\gamma'\geq-i\widetilde\sigma$, and $\gamma$ has
positive partial transpose, as a separable density operator
should. We have made this explicit, because it shows that it may
be interesting to see how much ``space'' there is between $\gamma$
and $-i\sigma$ and $-i\widetilde\sigma$. This leads to the central
definition of this paper:

\begin{Def}We say that a real symmetric matrix $\gamma$ is a {\it
ppt-covariance}, if $\gamma+i\sigma\geq0$ and
$\gamma+i\widetilde\sigma\geq0$, and that it is {\it minimally
ppt}, if it is a ppt-covariance, and any ppt-covariance $\gamma'$
with $\gamma\geq\gamma'$ must be equal to $\gamma$.\end{Def}

Note that a minimally ppt matrix $\gamma$ is separable if and only
if it is a direct sum, i.e., if the corresponding state
factorizes. There is a rather effective criterion for deciding
whether a given ppt-covariance is even minimally ppt: First of
all, if there were any $\gamma'\leq\gamma$ with
$\gamma'\neq\gamma$, we can also choose $\gamma-\gamma'=\Delta$ to
be a rank one operator, i.e., a matrix of the form
$\Delta_{\alpha\beta}=\xi_\alpha\xi_\beta$. Then we have
$\gamma+i\sigma\geq\epsilon\Delta$ for sufficiently small positive
$\epsilon$ if and only if $\xi$ is in the support of the positive
operator $\gamma+i\sigma$. The same reasoning applies to
$\widetilde\sigma$, so that $\gamma$ is minimally ppt iff there is
no real vector $\xi$, which is in the support of both
$\gamma+i\sigma$ and $\gamma+i\widetilde\sigma$. Rephrasing this
in terms of the orthogonal complements of the supports, we get the
following characterization which we will use later on:

\begin{Lem}\label{L:min} Let $\gamma$ be a ppt-covariance, and let
${\cal N}$ and $\widetilde{\cal N}$ denote the null spaces of
$\gamma+i\sigma$ and $\gamma+i\widetilde\sigma$, respectively.
Then $\gamma$ is minimally ppt if and only if ${\Re\!e}{\cal N}$
and ${\Re\!e}\widetilde{\cal N}$ together span $X$.
\end{Lem}
This gives an effective procedure to find a minimally ppt
$\gamma'$ below a given $\gamma$: in each step one subtracts the
largest admissible multiple of a rank one operator with vector
$\xi$ orthogonal to the span of ${\Re\!e}{\cal N}$ and
${\Re\!e}\widetilde{\cal N}$, which is then in the supports of
$\gamma+i\sigma$ and $\gamma+i\widetilde\sigma$. In every step
this will either increase ${\cal N}$ or $\widetilde{\cal N}$, so
that a minimally ppt covariance matrix is reached after a finite
number of steps.
\section{The $1\times N$ case}
This section is devoted to the proof that, for Gaussian states of
$1\times N$ systems, ppt implies separability. It is clear from
the previous section that this is equivalent to saying that every
minimally ppt covariance matrix is block diagonal, i.e., belongs
to a product state. So throughout this section we assume that
$\gamma$ is a minimally ppt covariance matrix.

As a first step we get rid of irrelevant pure state factors in the
following sense: Suppose that the two null spaces have a
non-trivial intersection, i.e., there is a $\Phi\neq0$ with
$\Phi\in{\cal N}\cap\widetilde{\cal N}$. Then
$(\sigma-\widetilde\sigma)\Phi=(i\gamma-i\gamma)\Phi=0$, so $\Phi$
has non-zero components only in Bob's part of the system. So let
$X_{C}$ denote the subspace of $X_B$ spanned by real and imaginary
part of $\Phi$. Then the restriction of the state to the subsystem
$C$ satisfies the pure state condition (its covariance matrix
$\gamma_C+i\sigma_C$ has a null vector by construction). It
follows that the density matrix factorizes: $\rho_{A,B\setminus
C,C}=\rho_{A,B\setminus C}\otimes\rho_C$, where $\rho_C$ is a pure
state. (This conclusion can also be obtained purely on the level
of covariance matrices, by introducing in $X_B$ a basis of
canonical variables containing a canonical basis of $X_C$).
Clearly, the separability of such a state is equivalent to the
separability of $\rho$, and the covariance matrix restricted to
$X_A\oplus X_{B\setminus C}$ is again minimally ppt. Hence we have
reduced the problem to the analogous one for the smaller space
$X_A\oplus X_{B\setminus C}$.

We may therefore assume without loss of generality that the null
spaces ${\cal N}$ and $\widetilde{\cal N}$ have trivial
intersection. This means that we proceed by contradiction, since
we want to prove ultimately that the state is a product of
``irrelevant pure state factors''.

Now let $0\neq\Phi\in{\cal N}$ and
$0\neq\widetilde\Phi\in\widetilde{\cal N}$. Then because $\gamma$
is hermitian, we have
$\langle\widetilde\Phi,\gamma\Phi\rangle=\langle\gamma\widetilde\Phi,\Phi\rangle$.
Using the null space conditions and the skew hermiticity of
$\sigma$, we can rewrite this as
\begin{equation}\label{s-tws}
   \langle\widetilde\Phi,(\sigma-\widetilde\sigma)\Phi\rangle=0\;.
\end{equation}
Now the vector $(\sigma-\widetilde\sigma)\Phi$ must be nonzero,
since otherwise we would have $\Phi\in{\cal N}\cap\widetilde{\cal
N}$. This is a condition on the $X_A$-components $\Phi_A$ of
$\Phi$, since $\sigma$ and $\widetilde\sigma$ differ only on that
two-dimensional subspace. By the same token the $X_A$-component
$\widetilde\Phi_A$ of $\widetilde\Phi$ must be non-zero. Hence all
vectors $(\sigma-\widetilde\sigma)\Phi$ lie in the one dimensional
subspace of $\Cx X_A$ orthogonal to $\widetilde\Phi_A\neq0$. The
proportionality constant is thus a linear functional on ${\cal N}$
vanishing only for $\Phi=0$, which means that ${\cal N}$ must be
one dimensional. By symmetry $\dim\widetilde{\cal N}=1$. By
Proposition~\ref{L:min} the spaces ${\Re\!e}{\cal N}$ and
${\Re\!e}\widetilde{\cal N}$ together span $X$, and since they are
two dimensional, it follows that $\dim X\leq4$, i.e., we can have
at most a $1\times1$ system. For such systems our claim has been
shown by Simon \cite{Simon}, and is hence proved.
\section{$2\times2$ bound entangled states}
It was already mentioned in the introduction that numerical
examples of minimally ppt covariances, which do not split into
$\gamma_A\oplus\gamma_B$ are easily generated by the subtraction
method. In contrast, the subtraction method for $1\times N$
systems always ends up at a block diagonal $\gamma$. This is
rather striking, but not really conclusive, because the numerical
determination of the null space of a matrix which may have small
eigenvalues may depend critically on rounding errors. We have
therefore prepared the following all integer $2\times2$ example
$\gamma$:
\begin{equation}\label{excov}
\gamma=\left(\begin{array}{cccccccc}
  2& 0& 0& 0& 1& 0& 0& 0\\
  0& 1& 0& 0& 0& 0& 0& -1\\
  0& 0& 2& 0& 0& 0& -1& 0\\
  0& 0& 0& 1& 0& -1& 0& 0\\
  1& 0& 0& 0& 2& 0& 0& 0\\
  0& 0& 0& -1& 0& 4& 0& 0\\
  0& 0& -1& 0& 0& 0& 2& 0\\
  0& -1& 0& 0& 0& 0& 0& 4
  \end{array}\right)\;.
\end{equation}
The key to getting simple examples is symmetry, which in turn
simplifies the verification of the basic properties.  The most
important symmetry in the example is the multiplication operator
$S$ with diagonal matrix elements $(1, 1, -1, -1, 1, -1, -1, 1)$.
It satisfies $S\sigma+\widetilde\sigma S=0$, and $S\gamma=\gamma
S$. Consequently, $\gamma+i\sigma$ and
$\gamma-i\widetilde\sigma=S(\gamma+i\sigma)S$ are unitarily
equivalent, so it suffices to check the positivity and to compute
the null space of $\gamma+i\sigma$. We note in passing that this
unitary equivalence is not necessary for bound entangled
Gaussians, since generically the spectra of $\gamma+i\sigma$ and
$\gamma+i\widetilde\sigma$ are different.

Further unitaries commuting with the covariance matrix
(\ref{excov}) are the multiplication operator $C$ with diagonal
matrix elements $(1, -1, 1, -1, 1, -1, 1,- 1)$, and the skew
symmetric operator $R$ with $R_{13}=R_{24}=R_{75}=R_{86}=1$, and
zero remaining entries. All these operators have square $\pm{\bf
1}$, and commute with each other and the symplectic forms up to
signs. Therefore, if we start with a generic vector
$\Omega_{1}\in{\cal N}$, the application of $R,C,S$ and products
of these operators yields eight vectors $\Omega_i$, which form a
basis of $\Cx^8$. Since these vectors lie in either ${\cal
N},\widetilde{\cal N}$ or their complex conjugates we know how
$\gamma$ acts on them and the covariance matrix is thus determined
by $\gamma = \Lambda \Omega^{-1}$, where $\Lambda, \Omega$ denote
the matrices consisting of column vectors $\Lambda_k = \gamma
\Omega_k$.  The above $\gamma$ is generated in this manner from
\begin{equation}\label{xi0}
  \Omega_{1}=(-1, i, 2, -3 i, 1, -i, 1, 0)\;.
\end{equation}
Then the condition of Proposition~\ref{L:min} is satisfied by
construction, and we only have to verify that
$\gamma+i\sigma\geq0$, which is again simplified by this operator
commuting with $R$. Explicitly, we get the eigenvalues
$0,3-\sqrt3,3,3+\sqrt3$, each with multiplicity $2$.

Generalizing this example we can construct a five parameter family
of bound entangled Gaussian states commuting with $R,S$ and $C$ in
the same manner as above. We start with a generic vector
\begin{equation}\label{xi0general}
  \Omega_{1}=(-a, i b, c, -i d, e, -i f, 1, 0)\;,
  \quad \;a, b,\ldots,f >0.
\end{equation}
Then $\gamma$ being real and symmetric requires $d=(b c + f)/a$
and from the characteristic function of $\langle\Omega_k
,\Lambda_l+i\sigma\Omega_l\rangle$ we obtain that
$\gamma+i\sigma\geq0$ iff $a \leq c e$, where equality is ruled
out since this would be equivalent to
 $\det(\Omega)= 0$.

States obtained from (\ref{xi0general}) are all of a non block
diagonal form similar to (\ref{excov}), and are hence bound
entangled.
\section*{Acknowledgement}
Funding by the European Union project EQUIP (contract
IST-1999-11053) and financial support from the DFG (Bonn) is
gratefully acknowledged.

\end{document}